\documentclass[twocolumn, twocolappendix]{aastex631}
\usepackage{amsmath}
\usepackage{mathrsfs}
\usepackage{graphicx}
\usepackage{xcolor}
\usepackage{dcolumn}
\usepackage{bm}
\usepackage{amssymb}
\usepackage{amsmath}
\usepackage{verbatim}
\usepackage[utf8]{inputenc}
\usepackage{tikz,hyperref}
\usepackage{dcolumn}
\usepackage{multirow}

\shortauthors{Yang, Bi, Wang, \& Yin}

\begin{document}

\title{Evolution of Compact Stellar Systems in Ultralight Dark Matter Halos: Dependence on Stellar and Dark Matter Parameters}

\author[0009-0005-5375-9437]{Yu-Ming Yang}
\email{yangyuming@ihep.ac.cn}
\affiliation{State Key Laboratory of Particle Astrophysics, Institute of High Energy Physics, Chinese Academy of Sciences, Beijing 100049, China}
\affiliation{School of Physical Sciences, University of Chinese Academy of Sciences, Beijing 100049, China }

\author[0000-0002-5334-9754]{Xiao-Jun Bi}
\email{bixj@ihep.ac.cn}
\affiliation{State Key Laboratory of Particle Astrophysics, Institute of High Energy Physics, Chinese Academy of Sciences, Beijing 100049, China}
\affiliation{School of Physical Sciences, University of Chinese Academy of Sciences, Beijing 100049, China }

\author[0000-0001-8713-0366]{Long Wang}
\email{wanglong8@sysu.edu.cn}
\affiliation{School of Physics and Astronomy, Sun Yat-sen University, Zhuhai 519082, China}

\author[0000-0001-6514-5196]{Peng-Fei Yin}
\email{yinpf@ihep.ac.cn}
\affiliation{State Key Laboratory of Particle Astrophysics, Institute of High Energy Physics, Chinese Academy of Sciences, Beijing 100049, China}

\begin{abstract}
Compact stellar systems are often used to place stringent constraints on the particle mass of ultralight dark matter (ULDM), as the heating effect induced by wave interference can drive system expansion, potentially bringing them into tension with observations. In a recent study, we pointed out that internal two-body relaxation in these stellar systems may have a significant impact on their evolution in ULDM halos, an effect overlooked in previous studies. Here, we further investigate the influence of stellar metallicity, the Milky Way's tidal field, and the ULDM particle mass on the long-term fate of compact stellar populations. We find that metal-richer systems are generally more resistant to disruption. The tidal field of the Milky Way, by altering the orbital motion of the stellar systems within host ULDM halos, can significantly affect their stability. Furthermore, we find in our simulations that the heating effect becomes stronger with increasing ULDM particle mass when the system size is much smaller than the ULDM de Broglie wavelength $R_{\rm h} \ll \lambda_{\rm dB} $, in contrast to the $\lambda_{\rm dB}\lesssim R_{\rm h}$ case. These results highlight the complexity of the evolution of compact stellar systems in ULDM halos, and suggest that existing constraints derived from the systems, such as ultrafaint dwarf galaxies, may require careful revision.
\end{abstract}

\keywords{
\href{http://astrothesaurus.org/uat/353}{Dark matter (353)};
\href{http://astrothesaurus.org/uat/416}{Dwarf galaxies(416)};
\href{http://astrothesaurus.org/uat/1880}{Galaxy dark matter halos (1880)}
}

\section{Introduction}
The nature of dark matter remains one of the most profound mysteries in modern physics \citep{1970ApJ...159..379R, Bertone:2016nfn, Cirelli:2024ssz}. In the standard cosmological paradigm, dark matter is modeled as cold dark matter (CDM), a non-relativistic, collisionless, and dissipationless component \citep{Vogelsberger:2019ynw}. However, the lack of definitive detection signals, together with several challenges faced by CDM on small astrophysical scales \citep{Bullock_2017, Tulin:2017ara, Sales:2022ich}, has motivated growing interest in alternative dark matter models. Among these, ultralight dark matter (ULDM) \citep{Hu_2000, Peebles_2000, Hui_2017, Hui_2021, Ferreira_2021}, composed of bosons with masses $m_a \sim 10^{-22}\,\mathrm{eV}$, has emerged as a particularly promising candidate. At typical galactic velocities, such particles possess de Broglie wavelengths on kiloparsec scales, potentially alleviating several small-scale tensions of CDM while preserving its success on large scales.

However, in recent years, increasingly stringent constraints have pushed the allowed ULDM particle mass to higher values, rendering its astrophysical behavior progressively more CDM-like and thereby reducing its ability to resolve the small-scale challenges that originally motivated the model. These constraints arise from both high-redshift observations, such as the Lyman-$\alpha$ forest \citep{Ir_i__2017, Rogers_2021}, and local dwarf galaxies, including those derived from Jeans analyses \citep{Hayashi_2021, Zimmermann:2024xvd} and dynamical-heating arguments \citep{Marsh:2018zyw, Dalal:2022rmp, Teodori:2025rul}. The latter are directly relevant to this work. The heating effect originates from gravitational-potential fluctuations induced by wave interference in ULDM, which transfer energy to stellar systems embedded within ULDM halos \citep{Bar_Or_2019, El_Zant_2019, Dutta_Chowdhury_2023, Yang:2024vgw, Yang:2024ixt, zhao2025semianalyticmodeleffectsfuzzy}. As a result, these systems expand and develop larger velocity dispersions over time, potentially becoming inconsistent with observations if the heating is sufficiently strong.

Specifically, \citet{Dalal:2022rmp} derived a lower bound of $m_a > 3 \times 10^{-19}\,\mathrm{eV}$ using the ultrafaint dwarf galaxies (UFDs) Segue 1 and Segue 2, which have half-light radii of $R_\mathrm{h} \sim 24.2\,\mathrm{pc}$ and $\sim 40.5\,\mathrm{pc}$, respectively. By examining larger dwarf galaxies such as Fornax ($R_\mathrm{h} \sim 0.7\,\mathrm{kpc}$), \citet{Teodori:2025rul} obtained a weaker constraint of $m_a \sim 5\times 10^{-21}\,\mathrm{eV}$. More recently, \citet{Yang:2025bae} showed that tidal stripping by the Milky Way (MW) can remove high-energy ULDM states, suppress wave interference, and consequently reduce the associated dynamical heating experienced by stellar systems. Using Fornax as an example, they demonstrated that this effect could relax previous constraints, thereby allowing consistency with $m_a \sim 10^{-22}\,\mathrm{eV}$. 

However, \citet{May:2025ppj} argued that such tidal suppression is effective only for sufficiently small ULDM particle masses \citep{caputo2026influencetidesselfgravityultralight}. For larger $m_a$, where the soliton size (comparable to the de Broglie wavelength $\lambda_{\rm dB}$) is much smaller than the tidal radius, multiple high-energy states survive tidal stripping and the heating effect remains significant. Consequently, compact systems such as Segue~1 can still exclude this high-mass regime even when tidal stripping is taken into account. For smaller ULDM masses, they further pointed out that MW satellite galaxies typically evolve in isolation for $\sim 1-2\,\mathrm{Gyr}$ before infall. During this phase, a system as compact as Segue~1 would be disrupted by ULDM-induced heating within $\sim 0.1\,\mathrm{Gyr}$, making its survival unlikely.

Nevertheless, \citet{Eberhardt:2025lbx} pointed out that when the stellar system's size $R_{\rm h}$ is much smaller than $\lambda_{\rm dB}$, the heating effect is suppressed by a factor of $R_\mathrm{h} / \lambda_{\rm dB}$. Consequently, for sufficiently small $m_a$, systems such as Segue~1 may survive even during their isolated evolution outside the MW, potentially weakening the corresponding ULDM constraints. In a recent study \citep{Yang:2026wdk}, we investigated this scenario using numerical simulations. We found that if the initial stellar system was substantially more massive and compact than observed today, stellar two-body relaxation \citep{1987degc.book.....S, Meylan_1997, 2008gady.book.....B} becomes important and competes with ULDM-induced heating. Depending on the initial conditions, the system may undergo core collapse, disruption, or reach a quasi-stationary state. Within this framework, Segue~1 can be interpreted as a system on the verge of disruption even for $m_a \sim 10^{-22}\,\mathrm{eV}$. We further showed that including the stellar initial mass function (IMF) and stellar evolution (SE) enriches the dynamical evolution of the system, while still allowing the formation of Segue~1-like systems.

In this work, we further explore the impact of various physical parameters on the evolution of these systems. We incorporate the effects of ULDM into direct $N$-body simulations of stellar systems through a first-order tidal tensor approximation. The $N$-body simulations are carried out using PeTar \citep{Wang_2020}, which includes an SE package based on the framework of SSE/BSE \citep{Hurley_2000, Hurley_2002}. We find that systems with higher stellar metallicity are generally more resistant to disruption. Furthermore, we show that the heating effect is primarily driven by the tidal influence of the soliton on the stellar system. We also find that the orbital motion of the stellar system differs between an isolated ULDM halo and a ULDM subhalo embedded within the MW's tidal field, thereby changing the frequency of passages through the vicinity of the soliton and consequently affecting the internal evolution of the stellar system. Meanwhile, our simulations show the trend that the heating effect becomes stronger for larger ULDM particle masses, causing the stellar system to dissolve more rapidly. This trend is opposite to that found in the regime where the size of the stellar system exceeds the de Broglie wavelength of the ULDM field.

The paper is organized as follows. In Section~\ref{Sec2}, we describe the simulation setup. The simulation results and their analysis are presented in Section~\ref{Sec3}. Finally, we summarize our main findings in Section~\ref{Sec4}.
\section{Simulation Setup\label{Sec2}}.

\begin{table*}[htbp]
    \centering
    \caption{Summary of the simulation sets in this work}
    \begin{tabular}{ccccc}
    \hline
    \hline
    &$Z$&$m_{22}$&MW tidal after 3 Gyr&Stellar system motion\\
    \hline
    S0&$10^{-4}$ &$1$&Yes&Free \\
    S1&$10^{-2}$ &$1$&Yes&Free \\
    S2&No SE &$1$&Yes&Free \\
    S3&No SE, IMF &$1$&Yes&Free \\
    S4&$10^{-4}$ &$1$&No&Free \\
    \hline
    S0-soliton&$10^{-4}$ &1&No&Same as S0 \\
    S4-soliton&$10^{-4}$ &1&No&Same as S4 \\
    \hline
    S-fixed-1&$10^{-4}$ &$1$&No&Fixed \\
    S-fixed-1.5&$10^{-4}$ &$1.5$&No&Fixed \\
    S-fixed-2&$10^{-4}$ &$2$&No&Fixed \\
    S-fixed-3&$10^{-4}$ &$3$&No&Fixed \\
    \hline
    \end{tabular}
    \label{Tab1}
    \tablecomments{Column 1 lists the simulation labels. Column 2 gives the adopted stellar metallicity. Column 3 specifies the ULDM particle mass $m_{22}\equiv m_a/10^{-22}\,\mathrm{eV}$. Column 4 indicates whether MW tidal stripping of the ULDM halo is included after $3\,\mathrm{Gyr}$. Column 5 indicates whether the stellar system evolves freely within the ULDM halo or is fixed at the halo center of mass. SE is switched off in the S2 simulation, while both SE and the IMF are neglected in the S3 simulation. The simulations S0-soliton and S4-soliton employ a static soliton potential, with the tidal tensor recorded along the corresponding stellar-system orbits.}
\end{table*}

We adopt the same two-stage simulation strategy as described in \cite{Yang:2026wdk}. Here, we provide only a summary, while a detailed description can be found in \cite{Yang:2026wdk}.

First, we construct the initial wavefunction of the ULDM halo using the eigenstate decomposition method \citep{Lin_2018, Yavetz_2022, Yang:2024trr}. To ensure that the halo reaches a quasi-steady state, the generated wavefunction is evolved in isolation for $1\,\mathrm{Gyr}$ before the main simulation begins. The stellar system is then treated as a point particle and initially placed at rest at the coordinate origin. The entire system is subsequently evolved self-consistently. For a subset of our simulations, we further consider the tidal stripping of the outer ULDM halo by the MW. The system is assumed to orbit within the MW potential following the orbit adopted in \cite{Yang:2026wdk}. The corresponding external tidal potential is incorporated into the ULDM evolution according to the prescription of \cite{Yang:2025bae}, thereby enabling a self-consistent treatment of MW-induced tidal stripping. During the entire evolution, we continuously record the ULDM tidal tensor, $\nabla\nabla V_{\rm ULDM}(t,\boldsymbol{x})$, evaluated at the position of the point particle.

The ULDM wavefunction is evolved using the pseudo-spectral method implemented in the PyUltraLight package \citep{Edwards_2018}, while the position and velocity of the point particle are updated using a fourth-order Runge-Kutta integrator. The wavefunction is evolved in a cubic box with a side length of $40\,\mathrm{kpc}$. Unless otherwise stated, a spatial resolution of $256^3$ and a timestep of $1\,\mathrm{Myr}$ are adopted for the wavefunction evolution, while the point particle orbit is integrated with a timestep of $0.1\,\mathrm{Myr}$.

In the second stage, we follow the internal evolution of the stellar system. The initial stellar distribution is assumed to follow a Plummer profile \citep{1911MNRAS..71..460P} $\rho_\star(r)=(3M_\star/4\pi R_\mathrm{h}^3)(1+r^2/R_\mathrm{h}^2)^{-5/2}$. Approximately equilibrium initial conditions are generated using McLuster \citep{K_pper_2011}, and the subsequent evolution is carried out with the direct $N$-body code PeTar \citep{Wang_2020}. In addition to the gravitational forces exerted by other stars, we incorporate the effect of ULDM through the tidal acceleration
\begin{equation}
    \Delta\boldsymbol{a}(\boldsymbol{r},\boldsymbol{r}_c)\simeq -(\boldsymbol{r}-\boldsymbol{r}_c)\cdot\nabla\nabla V_\text{ULDM}(\boldsymbol{r}_c),
    \label{tidal_tensor}
\end{equation}
which is added to the acceleration of each stellar particle at every timestep. Here, $\boldsymbol{r}$ and $\boldsymbol{r}_c$ denote the position of an individual stellar particle and the center of mass of the stellar system, respectively. Therefore, $\Delta \boldsymbol{a}$ represents the acceleration of a stellar particle relative to the center of mass of the stellar system induced by the ULDM tidal field.

In our previous work \citep{Yang:2026wdk}, we investigated how the number of stellar particles $N$, the initial stellar half-light radius $R_{\rm h}$, and the high-mass-end slope of the IMF affect the evolution of the system. In the present study, we therefore fix the first two quantities at $N=5\times 10^4$ and $R_h=5\,\mathrm{pc}$, and adopt an IMF of the form $\xi(m)\propto m^{-1.3}$ for $0.08\, M_\odot < m \leq 0.5\, M_\odot$ and $\xi(m)\propto m^{-2.4}$ for $m > 0.5\, M_\odot$, with an upper mass cutoff of $100\, M_\odot$. We conduct 11 simulation runs, with the main parameters and brief descriptions summarized in Table~\ref{Tab1}. The motivation for each simulation setup, along with the corresponding implementation details, will be discussed in the following sections.

\section{Results\label{Sec3}}
\subsection{Effects of stellar system internal parameters}
In our previous work, we showed that black holes (BHs) with masses of $\mathcal{O}(10)\, M_\odot$, formed through SE, can transfer low-mass stars toward the outer regions of the system via mass segregation. As a result, the stellar system becomes more susceptible to disruption under the influence of ULDM fluctuations. Since stellar metallicity is a key parameter governing SE, it may profoundly influence the system's long-term evolution.

To investigate the role of stellar metallicity, we adopt metallicities of $Z=10^{-4}$ and $Z=10^{-2}$ for the simulation sets S0 and S1, respectively. In the S2 set, SE is completely switched off. In the S3 set, we additionally neglect the IMF by assigning equal mass to all stellar particles while preserving the total stellar mass. To ensure that the ULDM influence remains identical across the four simulation sets, we adopt the same tidal tensor for all runs. This tidal tensor is taken from our previous work \citep{Yang:2025bae}, in which the ULDM particle mass was set to $m_{22}\equiv m_a/10^{-22}\,\mathrm{eV}=1$. In that setup, the combined ULDM-stellar system was first evolved in isolation for $3\,\mathrm{Gyr}$, followed by an additional $7\,\mathrm{Gyr}$ evolution within the MW's tidal field. Therefore, the S0 simulation is identical to the run presented in \cite{Yang:2026wdk} that reproduces a Segue 1-like system, and is adopted here as a benchmark simulation.

\begin{figure}[htbp] 
    \centering    
    \includegraphics[width=\linewidth]{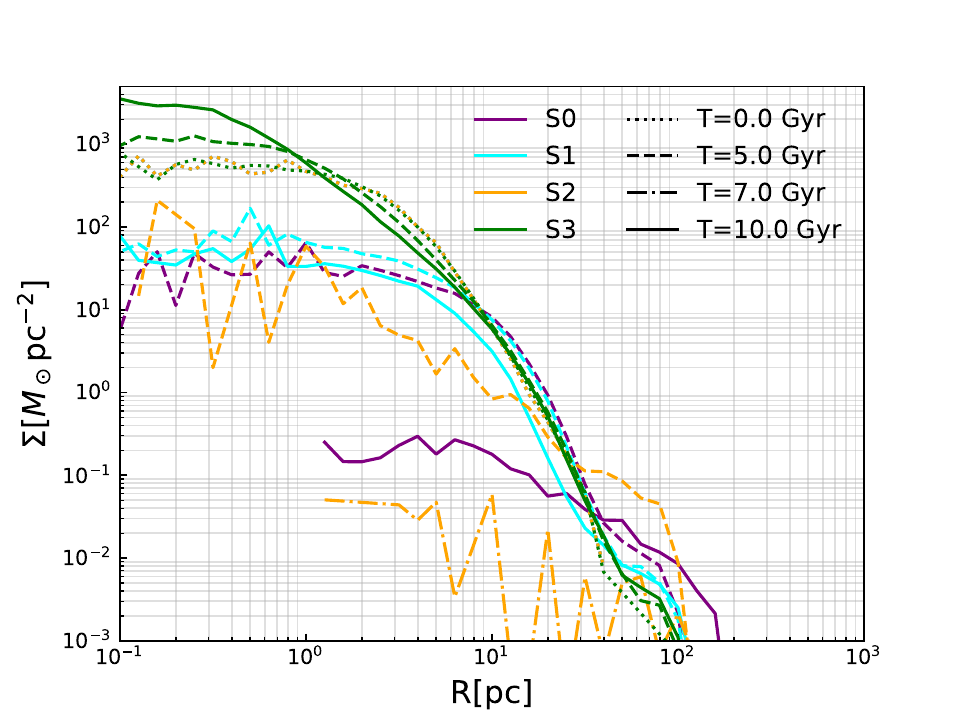}
    \includegraphics[width=\linewidth]{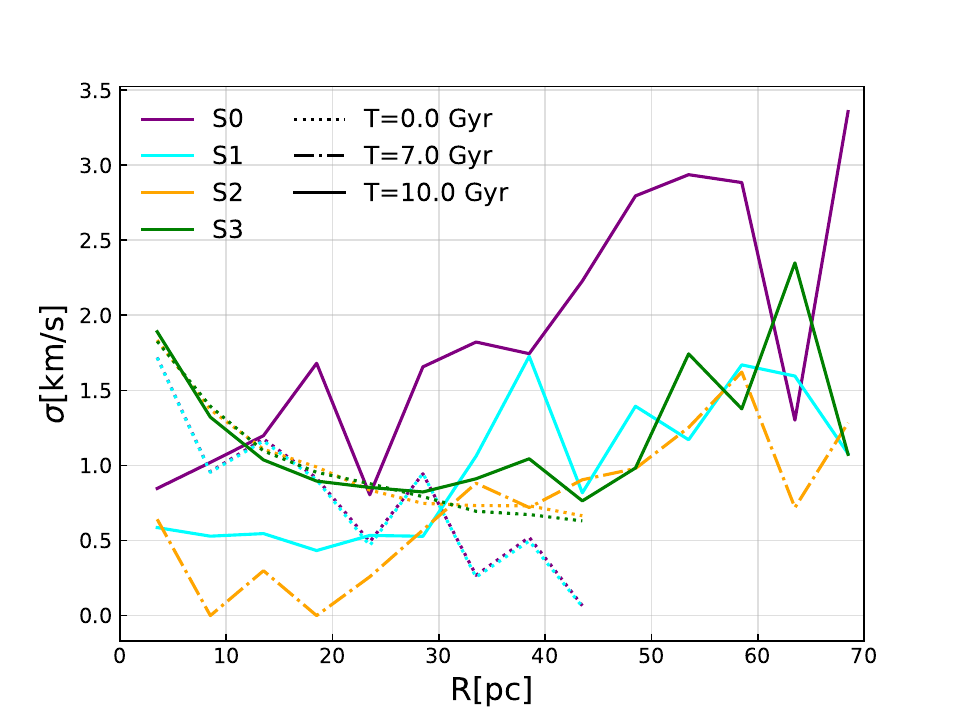}
    \caption{Upper panel: Projected stellar surface density profiles along the $z$-direction at different snapshots for the simulation sets S0, S1, S2, and S3. Colors and line styles denote the simulation sets and evolutionary times, respectively. Lower panel: Corresponding dispersion profiles of the $z$-component of the stellar velocities.}
    \label{Sigma_velocity_S0123}
\end{figure} 

The projected stellar surface density profiles along the z-direction at several snapshots for the four simulation sets are shown in the upper panel of Figure~\ref{Sigma_velocity_S0123}. Different colors and line styles distinguish the simulation sets and evolutionary times, as indicated in the figure. Since S0, S1, and S2 share identical initial conditions, their initial stellar distributions completely overlap; S3 differs slightly owing to the absence of an IMF.

\begin{figure}[htbp]
    \centering    
    \includegraphics[width=\linewidth]{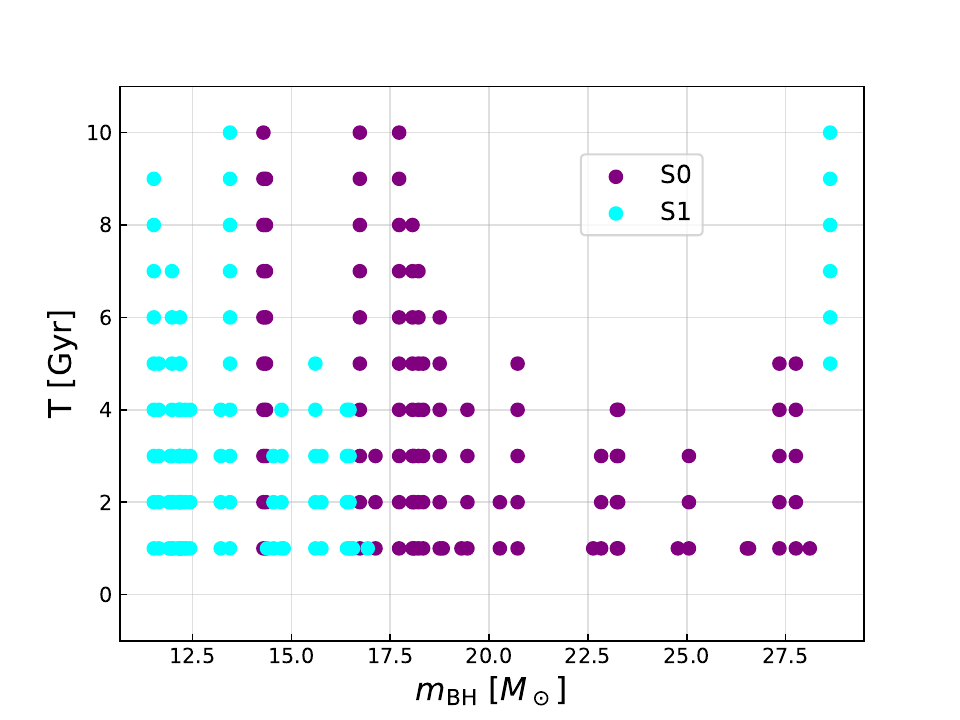}
    \caption{BHs within $60\,\mathrm{pc}$ in the simulations S0 and S1 at different snapshots. Each column represents an individual BH.}
    \label{BH_S01}
\end{figure}

It can be clearly seen that the S0 system is close to disruption by $10\,\mathrm{Gyr}$, whereas S1 remains relatively stable. The origin of this difference can be understood from Figure~\ref{BH_S01}, in which we show the masses of all BHs within $60\,\mathrm{pc}$ in S0 (purple) and S1 (cyan) at different epochs, with each point representing an individual BH. Initially, no BHs are present, and all stars are on the main sequence. During the first few tens of Myr, stellar evolution produces a population of BHs in both systems. Thereafter, the number of BHs gradually decreases as the systems are continuously heated by ULDM. Our analysis begins at $1\,\mathrm{Gyr}$, when the initial phase of BH formation has already been completed. At that time, S0 and S1 contain 26 and 28 BHs within $60\,\mathrm{pc}$, respectively. At any given epoch, the BHs in S1 are systematically less massive than those in S0. This is a natural consequence of the higher stellar metallicity adopted in S1, since metal-rich massive stars experience stronger stellar winds during their evolution and therefore lose more mass before core collapse \citep{Vink:2001cg, Eldridge_2006}. As a result, the mass segregation driven by BHs is weaker in S1 than in S0, thereby slowing down the disruption of the stellar system. In S1, a stochastic $\sim 28.6\, M_\odot$ BH appears at $\sim 5\,\mathrm{Gyr}$, formed through the merger of two lower-mass BHs. Although it is more massive than any BH in S0, it does not induce significant mass segregation in our simulations.

The S2 system is nearly disrupted by $7\,\mathrm{Gyr}$, which is even earlier than S0. This is because SE is switched off in S2, preventing mass loss from massive stars and thereby intensifying mass segregation. Massive stars efficiently transfer low-mass stars toward the outer regions of the system, thereby accelerating the disruption of the stellar system. Conversely, in S3, where all stellar particles have equal mass and SE is neglected, mass segregation is completely absent. As a result, this system does not undergo disruption and instead eventually experiences core collapse. This behavior is consistent with the phase diagram presented in \cite{Yang:2026wdk}, in which this system lies within the ``core collapse'' regime.

The lower panel of Figure~\ref{Sigma_velocity_S0123} shows the $z$-component velocity dispersion profiles at different epochs. In S0, S1, and S2, the stellar density in the central region gradually decreases during the evolution, shallowing the gravitational potential and reducing the velocity dispersion at small radii. In contrast, S3 remains nearly unchanged. At larger radii, ULDM heating raises the velocity dispersion in all four systems.

\begin{figure}[htbp]
    \centering    
    \includegraphics[width=\linewidth]{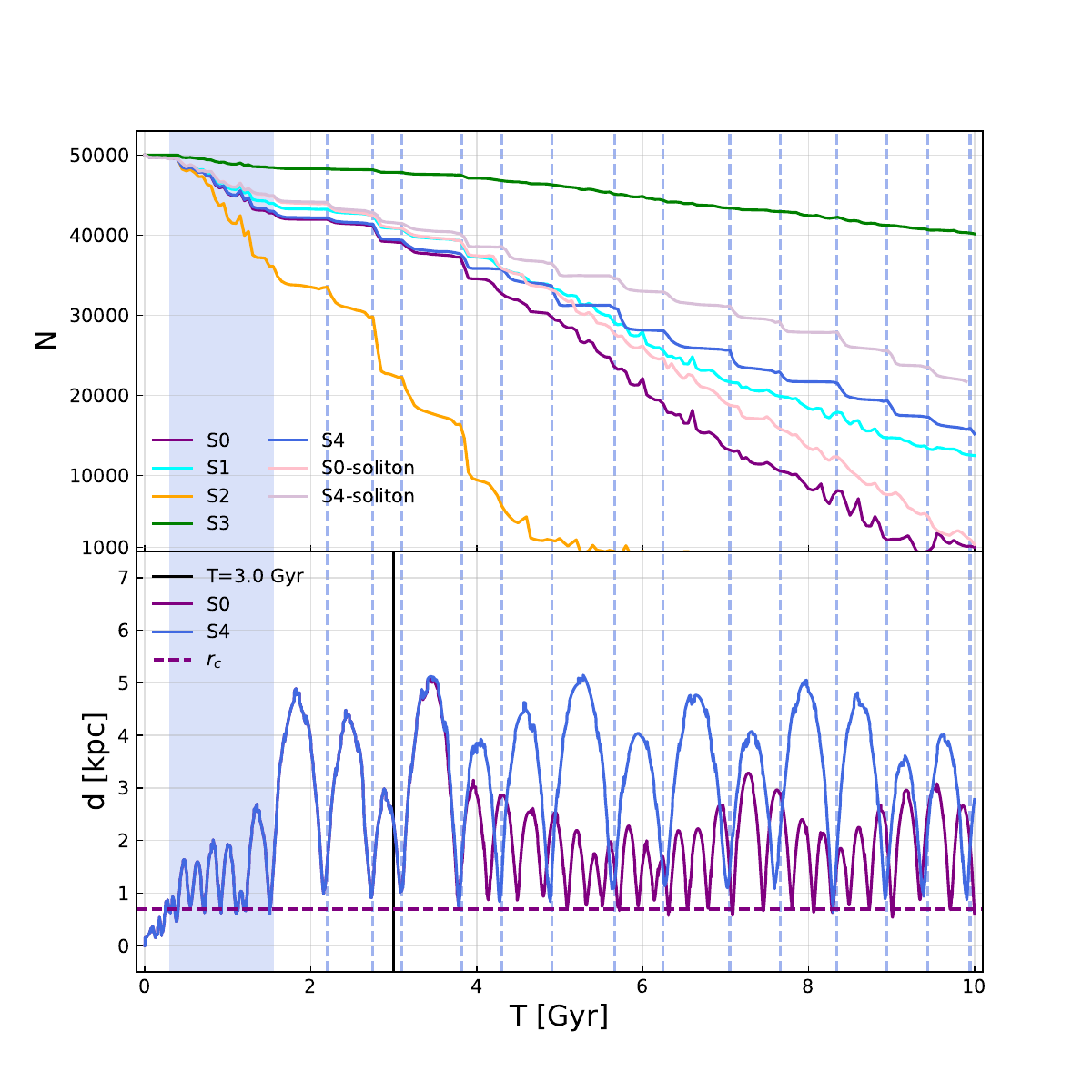}
    \caption{Upper panel: Temporal evolution of the number of stellar particles within $60\,\mathrm{pc}$ across the various simulation sets. Lower panel: Temporal evolution of the distance between the stellar system and the soliton center for S0 and S4 . The purple dashed line denotes the soliton radius, $r_\mathrm{c}$, defined as the radius at which the soliton density drops to half its central value. Blue shaded regions and blue dashed lines in both panels indicate epochs of rapid stellar-particle loss in the simulations S4 and S4-soliton.}
    \label{N_T_orbit}
\end{figure}

The upper panel of Figure~\ref{N_T_orbit} tracks the number of stellar particles within $60\,\mathrm{pc}$ as a function of time, further indicating the different disruption rates of the four systems.
By $10\,\mathrm{Gyr}$, S0, S1, and S3 retain approximately $\sim 10^3$, $\sim 1.2\times 10^4$, and $\sim 4\times 10^4$ stellar particles, respectively, whereas the S2 system already contains only $\sim 10^3$ particles by $5\,\mathrm{Gyr}$.

\subsection{Influence of orbital motion within ULDM halos}

In S0, originally designed to model a Segue~1-like system in our previous work \citep{Yang:2026wdk}, we consider a relatively realistic evolutionary path: the system first evolves in isolation for $3\,\mathrm{Gyr}$, then enters the Galactic tidal field, which strips the outer regions of the ULDM halo and thereby indirectly affects the embedded stellar system’s orbital motion. 
For comparison, in the S4 simulation set, we keep all other conditions identical to those in S0. The only difference is that, in the first-stage simulation, after the initial $3\,\mathrm{Gyr}$ evolution, S4 continues to evolve in isolation without ever subjection to the Galactic tidal field.

\begin{figure}[htbp]
    \centering    
    \includegraphics[width=\linewidth]{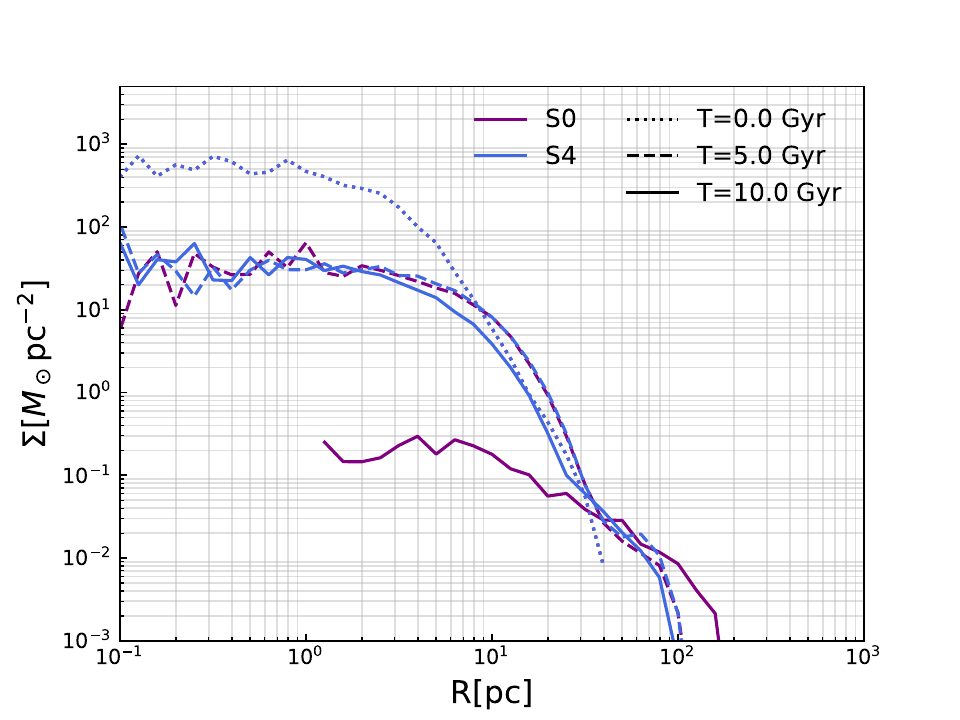}
    \caption{Projected stellar surface density profiles along the $z$-direction at different snapshots for S0 and S4.}
    \label{Sigma_S04}
\end{figure}

Figure~\ref{Sigma_S04} shows the temporal evolution of the stellar surface density distributions for S0 and S4. The S4 system remains relatively stable even at $10.0\,\mathrm{Gyr}$, whereas S0 approaches disruption. This difference is also evident in the upper panel of Figure~\ref{N_T_orbit}, where S4 retains $\sim 1.5\times 10^4$ stellar particles within $60\,\mathrm{pc}$ at $10\,\mathrm{Gyr}$, significantly more than S0. These findings indicate that, compared to S4, the inclusion of MW-induced tidal stripping of the outer ULDM halo in S0 enhances the ULDM heating effect and accelerates the disruption of the stellar system. This result is opposite to the conclusion obtained in previous studies within the $\lambda_\mathrm{dB}\lesssim R_\mathrm{h}$ regime, in which the Galactic tidal field suppresses dynamical heating by stripping high-energy states in the outer ULDM halo, thereby weakening interference fluctuations \citep{Yang:2025bae}.

This behaviour can be understood by examining the orbital motions of the stellar system within the ULDM halo. The lower panel of Figure~\ref{N_T_orbit} shows the evolution of the distance $d$ between the stellar system and the soliton center. Note that $d$ contains contributions both from the soliton's random walk \citep{Schive:2019rrw} and from the stellar system's motion within the ULDM halo. For simplicity, we hereafter refer to the stellar system's motion relative to the soliton center as its orbital motion. After the Galactic tidal field is introduced at $3\,\mathrm{Gyr}$ in S0, the progressive stripping of the outer halo causes the orbital motions in the two simulations to gradually diverge, with the difference becoming particularly pronounced after $\sim 4\,\mathrm{Gyr}$. Compared to S0, the stellar system in S4 exhibits a larger apocentric distance, a longer orbital period, and fewer passages through pericenter.

In S4, Figure~\ref{N_T_orbit} shows that the stellar particle number declines primarily when the system lies outside the soliton ($d > r_\mathrm{c}$) yet remains relatively close to it; these intervals are marked by the blue shaded region and blue dashed lines.
A similar behavior can also be seen in S0 before $4\,\mathrm{Gyr}$, where the loss of stellar particles is most pronounced near pericentric passages. After $4\,\mathrm{Gyr}$, the stellar system in S0 passes through pericenter much more frequently, leading to a nearly continuous decline in the number of bound stellar particles. Therefore, this result suggests that, in the $\lambda_\mathrm{dB}\gg R_\mathrm{h}$ regime, the dominant heating effect arises when the stellar system is displaced from the soliton center and experiences the strongly inhomogeneous gravitational potential of the soliton. This process is more accurately described as a form of tidal heating, as proposed by \cite{liu2026tidalheatingstellarclusters}.

\subsection{Dominance of soliton-induced heating}
In general, the heating effect results from both the soliton and the fluctuating granular structures generated by wave interference in the ULDM halo. In the following, we demonstrate that the soliton emerges as the dominant heating source in the $\lambda_\mathrm{dB} \gg R_\mathrm{h}$ regime, as shown in our simulations. 

To isolate the soliton's contribution, we construct the S0-soliton and S4-soliton simulations. The tidal tensors in these runs are generated as follows. We place a static soliton potential, without the surrounding NFW envelope, at the center of the simulation box. The soliton parameters are chosen to be identical to those adopted in the S0 and S4 simulations, respectively. The stellar system is forced to follow precisely the orbital trajectories displayed in the lower panel of Figure~\ref{N_T_orbit} for the S0 and S4 cases, and the tidal tensor is recorded along these trajectories. As a result, the S0-soliton and S4-soliton simulations retain only the soliton-induced heating while excising the influence of the fluctuating granules. 

\begin{figure}[htbp]
    \centering    
    \includegraphics[width=\linewidth]{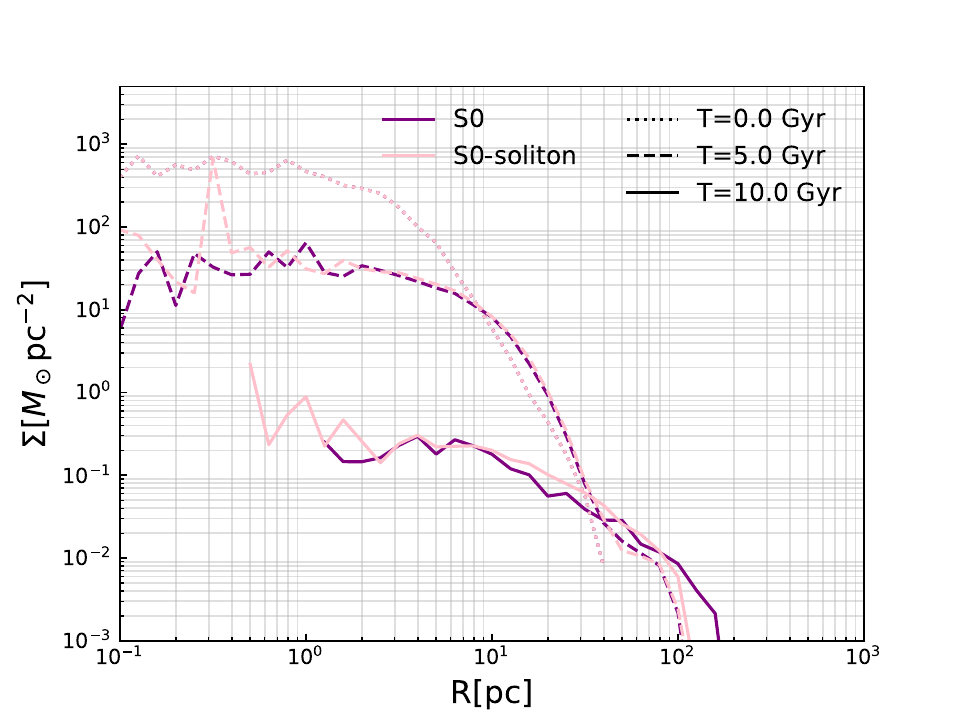}
    \includegraphics[width=\linewidth]{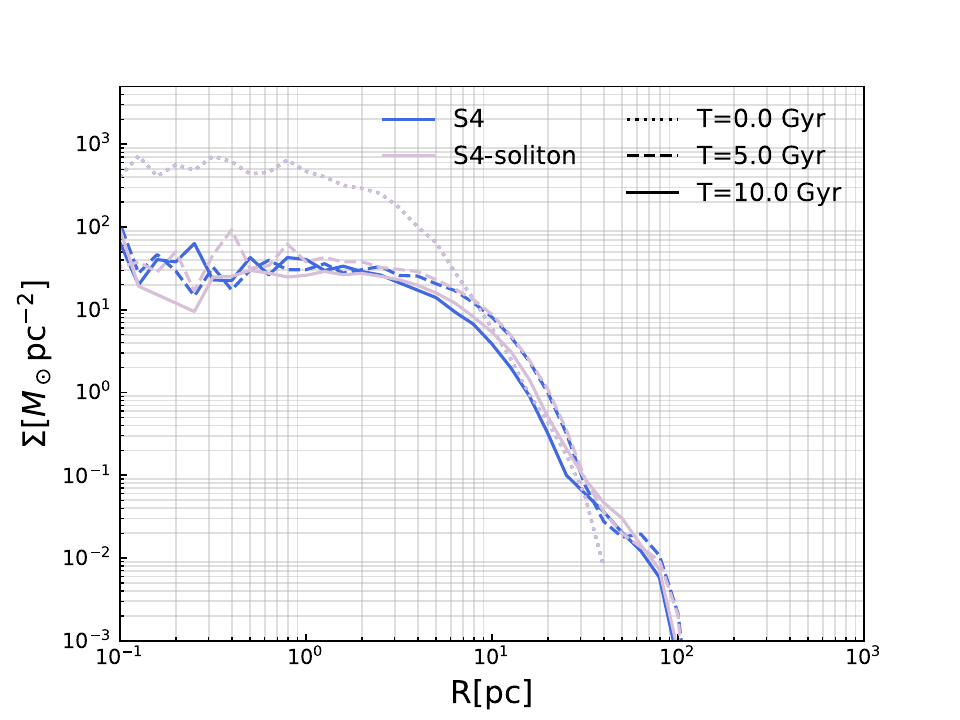}
    \caption{Upper panel: Projected stellar surface density profiles along the $z$-direction at different snapshots for the simulations S0 and S0-soliton. Lower panel: Same as the upper panel, but for the simulations S4 and S4-soliton.}
    \label{Sigma_S04soliton}
\end{figure}

In the upper panel of Figure~\ref{Sigma_S04soliton}, we compare the evolution of the stellar surface density profiles in the S0 and S0-soliton simulations. It is evident that the two runs produce remarkably similar results, demonstrating that the soliton is the dominant source of the heating effect. A closer inspection reveals that, over certain radial ranges, the stellar surface density in S0 is slightly lower than in S0-soliton. This difference is more directly visible in Figure~\ref{N_T_orbit}, where the number of stellar particles within $60\,\mathrm{pc}$ decreases more rapidly in S0 than in S0-soliton. These results indicate that the heating effect in S0 is slightly stronger than that from the soliton alone, with the additional contribution attributable to the fluctuating granular structures in the outer halo. A consistent picture emerges from the comparison between S4 and S4-soliton, shown in the lower panel of Figure~\ref{Sigma_S04soliton}.

The dominance of the soliton-driven heating can be understood qualitatively as follows. The characteristic amplitude of the tidal tensor induced by the soliton can be estimated as $GM_\mathrm{c}/r_\mathrm{c}^3\sim G\rho_\mathrm{c}$, whereas that generated by an individual granule is of order $GM_\mathrm{granule}/r_\mathrm{granule}^3\sim G\rho_\mathrm{granule}$, which lead to the acceleration difference within the stellar system as $\delta a\sim G\rho*R_h$. Since the density of the soliton $\rho_\mathrm{c}$ is generally much higher than that of the surrounding granules $\rho_\mathrm{granule}$, the tidal field produced by the soliton is correspondingly stronger. As a result, the impact of the soliton on the stellar system naturally dominates over that of the granular fluctuations, making the soliton the primary driver of the heating process.

\subsection{Dependence of heating strength on ULDM particle mass}
In the $\lambda_\mathrm{dB}\lesssim R_\mathrm{h}$ regime, the dynamical heating effect is found to decrease with increasing ULDM particle mass. However, in the $\lambda_\mathrm{dB}\gg R_\mathrm{h}$ regime, \cite{liu2026tidalheatingstellarclusters} recently showed that the heating effect becomes stronger with increasing ULDM particle mass, using massless tracer particles in simulations. This result is consistent with the theoretical analysis of \cite{Eberhardt:2025lbx}. Here, we revisit this issue by explicitly including the self-gravity of the stellar system.

To this end, we perform four simulation sets, designated S-fixed-1, S-fixed-1.5, S-fixed-2, and S-fixed-3. In the first-stage simulations, the stellar system is fixed at the center of mass of an isolated ULDM halo, and the tidal tensor at that location is recorded throughout the halo's evolution. The ULDM particle masses adopted in these runs are $m_{22}=1, 1.5, 2$, and $3$, respectively. For the three higher-mass cases, we increase the numerical resolution to $512^3$ in order to adequately resolve the correspondingly smaller de Broglie wavelengths.
 
\begin{figure}[htbp]
    \centering        \includegraphics[width=\linewidth]{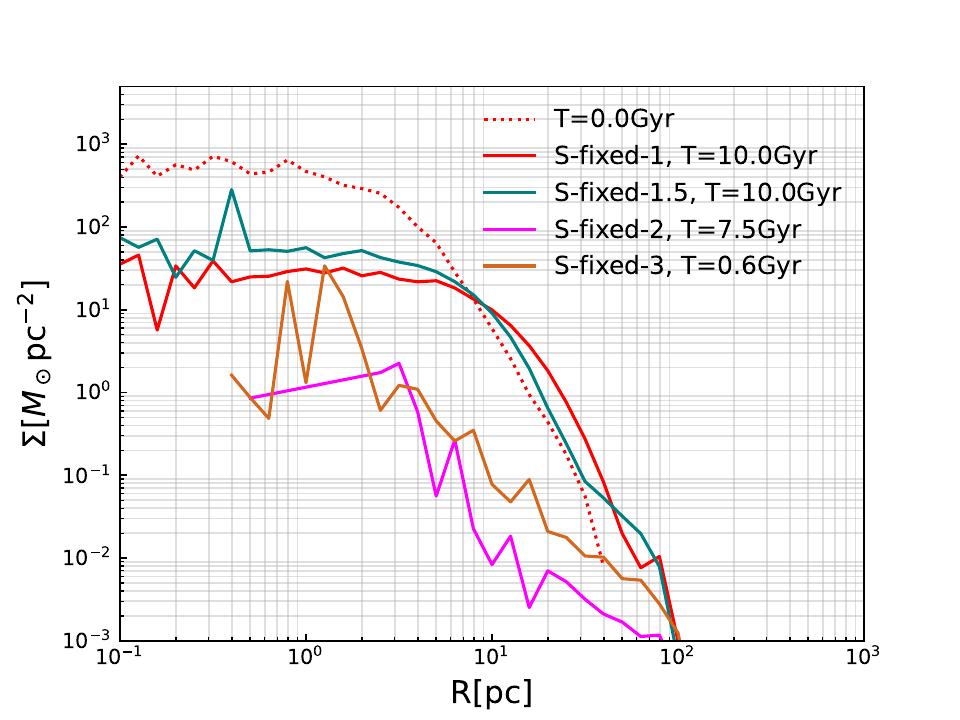}
    \caption{Projected stellar surface density profiles along the $z$-direction at different snapshots for the simulations S-fixed-1, S-fixed-1.5, S-fixed-2, and S-fixed-3.}
    \label{Sigma_S_fixed}
\end{figure}

The evolution of the stellar surface density profiles for these four simulation sets is shown in Figure~\ref{Sigma_S_fixed}. It can be seen that S-fixed-1.5 exhibits a density distribution at $10\,\mathrm{Gyr}$ similar to that of S0, whereas S-fixed-2 already approaches disruption by $7.5\,\mathrm{Gyr}$. The S-fixed-3 system disrupts even more rapidly, becoming nearly dissolved by $\sim 0.6\,\mathrm{Gyr}$. These results indicate a trend: the heating effect becomes stronger with increasing ULDM particle mass. However, it should be noted that the disruption time is subject to a degree of stochasticity as discussed below, and therefore does not provide a perfect measure of the heating strength.

\begin{figure}[htbp]
    \centering    
    \includegraphics[width=\linewidth]{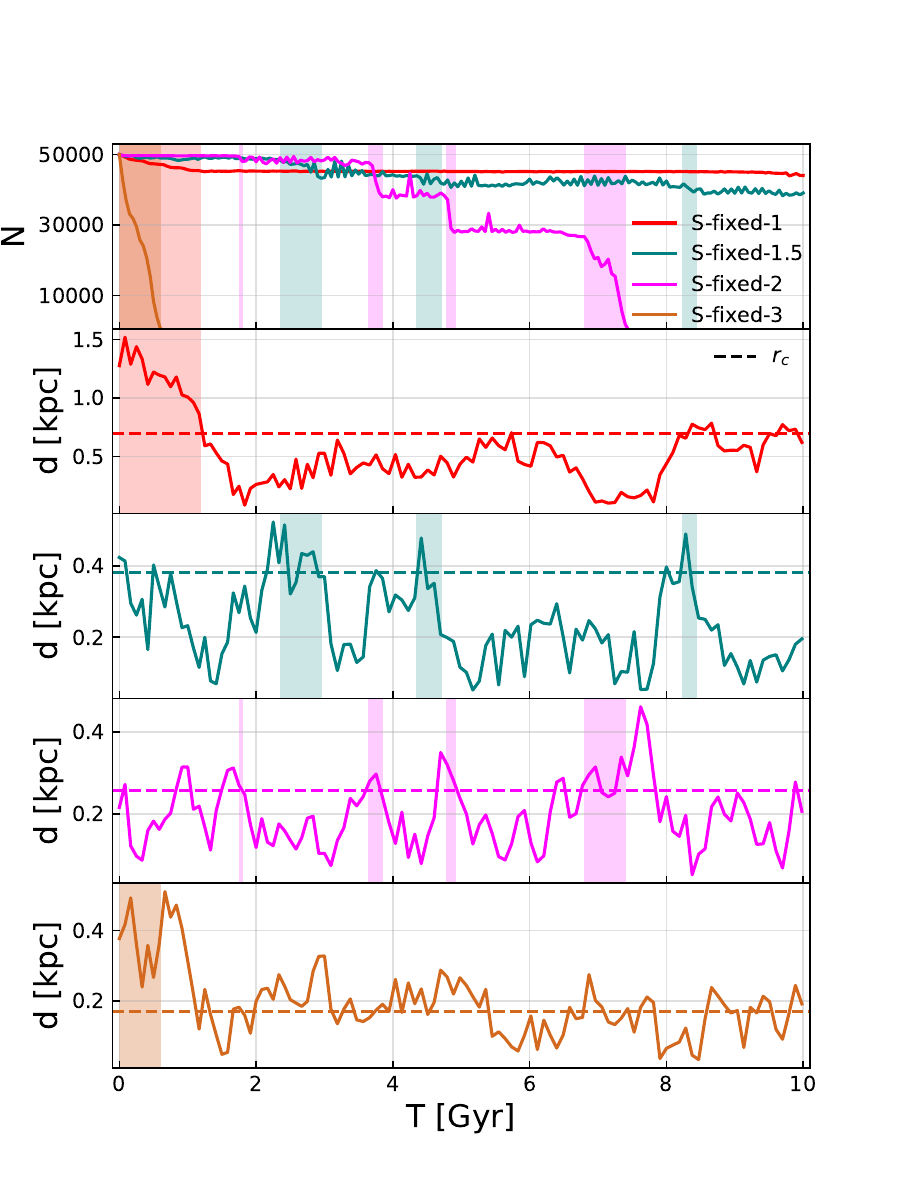}
    \caption{Top panel: Temporal evolution of the number of stellar particles within $60\,\mathrm{pc}$ for four simulation sets. Lower four panels: Temporal evolution of the distance between the stellar system and the soliton center for the simulations S-fixed-1, S-fixed-1.5, S-fixed-2, and S-fixed-3, respectively. The dashed lines denote the corresponding soliton radii. The shaded regions mark  epochs of rapid stellar-particle loss in the simulations.}
    \label{N_T_orbit_fixed}
\end{figure}

The top panel of Figure~\ref{N_T_orbit_fixed} traces the number of stellar particles within $60\,\mathrm{pc}$ over time for the four simulation sets. The four lower panels display the evolution of the distance between the stellar system (which is fixed at the center of mass of the ULDM halo) and the soliton center. Since the stellar system remains fixed throughout the simulations, the variation in this distance arises solely from the random walk of the soliton. The decrease in the number of stellar particles occurs in a step-like manner in all four simulations. Moreover, these decreases consistently take place when the stellar system is located at $d \gtrsim r_\mathrm{c}$, as highlighted by the shaded bands. In contrast, when the stellar system resides within the soliton ($d < r_\mathrm{c}$), the particle number remains nearly constant, indicating that ULDM has little impact on the internal evolution of the stellar system in this regime. This provides further support for our previous interpretation that the heating effect is primarily driven by the tidal influence of the soliton on the stellar system in our simulations.

On the other hand, because the soliton’s motion is stochastic, the times at which it migrates beyond $r_\mathrm{c}$ are also stochastic, rendering the timing of significant particle-loss events equally random. For example, we evolve each halo in isolation for $1\,\mathrm{Gyr}$ to reach a quasi-steady state before the main simulation begins. As a result, the soliton's position at the start of the simulation is itself random. In both S-fixed-1 and S-fixed-3, the soliton happens to be located relatively far from the halo center at the initial time, leading to an immediate decrease in particle number. This illustrates that the disruption time is subject to a degree of stochasticity and therefore does not provide a direct measure of the heating strength. Nevertheless, Figure~\ref{N_T_orbit_fixed} clearly shows that the magnitude of each particle-loss event increases with ULDM particle mass, providing a more robust indicator of the mass dependence of the heating effect.

The trend observed in our simulations, namely that the heating effect becomes stronger as the ULDM particle mass increases, can be understood as a direct consequence of our earlier result that the heating is primarily driven by the tidal effect of the soliton. Given the soliton core-halo relation \citep{Schive_2014}, which is incorporated into the construction of the halo initial conditions, the soliton mass and radius scale with the ULDM particle mass and halo mass according to $M_\mathrm{c}\propto m_{22}^{-1}M_{200}^{1/3}$ and $r_\mathrm{c}\propto m_{22}^{-1}M_{200}^{-1/3}$, respectively. The characteristic strength of the soliton tidal heating can then be estimated as $GM_\mathrm{c}/r_\mathrm{c}^3\propto m_{22}^2M_{200}^{4/3}$. This scaling implies that the soliton tidal field becomes stronger for larger ULDM particle masses, naturally leading to a stronger heating effect.

\section{Conclusions\label{Sec4}}
In this work, we investigate  the evolution of compact stellar systems embedded in ULDM halos using direct $N$-body simulations that incorporate ULDM effects through a first-order tidal-tensor approximation. Building upon our previous study, we explore the impact of stellar metallicity, MW tidal stripping, and the ULDM particle mass on the long-term evolution of these systems. We find that SE plays a crucial role in regulating disruption. Higher stellar metallicity leads to lower-mass BHs owing to stronger stellar winds, thereby weakening mass segregation and delaying disruption. In contrast, neglecting stellar evolution enhances mass segregation and accelerates the dissolution of the stellar system, while equal-mass systems would undergo core collapse rather than disruption.

We further find that the heating effect in the $\lambda_{\rm dB}\gg R_{\rm h}$ regime is dominated by the soliton, with the fluctuating granular structures providing only a subdominant contribution. The evolution of the stellar system is closely linked to its orbital motion relative to the soliton, and significant stellar-particle loss occurs primarily when the stellar system is displaced beyond the soliton radius. Consequently, MW-induced tidal stripping can significantly affect the evolution of the stellar system by modifying its orbit within the ULDM halo. Finally, we show that the heating effect becomes stronger with increasing ULDM particle mass in the $\lambda_{\rm dB}\gg R_{\rm h}$ regime, leading to more rapid stellar-particle loss.
These results reveal that the evolution of compact stellar systems in the $\lambda_{\rm dB}\gg R_{\rm h}$ regime differs qualitatively from that in the $\lambda_{\rm dB}\lesssim R_{\rm h}$ regime, emphasizing the necessity of treating the two regimes separately when deriving astrophysical constraints on ULDM.

\section*{acknowledgments}
This work is supported by the National Natural Science Foundation of China under grants No.12447105, No.12575113, No.12573041, and No.12233013. L.W. thanks the High-level Youth Talent Project (Provincial Financial Allocation) through the grant 2023HYSPT0706, the Fundamental Research Funds for the Central Universities, Sun Yat-sen University (2025QNPY04).

\bibliography{Refs}
\bibliographystyle{aasjournal}
\end{document}